\documentclass[twoside]{article}
\usepackage{qic}

\usepackage[english]{babel}
\usepackage{amsmath}
\usepackage{amsfonts}
\newcommand{\bra}[1]{\langle #1 |}
\newcommand{\ket}[1]{| #1 \rangle}
\newcommand{\braket}[1]{\langle #1 \rangle}

\renewcommand{\thefootnote}{\fnsymbol{footnote}}  

\begin{document}
\setlength{\textheight}{8.0truein}    

\runninghead{Improving Quantum Spatial Search in Two Dimensions}
            {Abhijith J. and A. Patel}

\normalsize\textlineskip
\thispagestyle{empty}
\setcounter{page}{1}

\copyrightheading{19}{7\&8}{2019}{555--574}

\vspace*{0.88truein}

\alphfootnote

\fpage{1}

\centerline{\bf
IMPROVING THE QUERY COMPLEXITY OF}
\vspace*{0.035truein}
\centerline{\bf QUANTUM SPATIAL SEARCH IN TWO DIMENSIONS}
\vspace*{0.37truein}
\centerline{\footnotesize
ABHIJITH J.\footnote{E-mail: abhijithj@iisc.ac.in} ~~and
APOORVA PATEL\footnote{E-mail: adpatel@iisc.ac.in}}
\vspace*{0.015truein}
\centerline{\footnotesize\it Centre for High Energy Physics,
Indian Institute of Science}
\baselineskip=10pt
\centerline{\footnotesize\it Bangalore 560012, India}
\vspace*{0.225truein}
\publisher{November 19, 2018}{May 17, 2019}

\vspace*{0.21truein}

\abstracts{
The question of whether quantum spatial search in two dimensions can be made optimal has long been an open problem.
We report progress towards its resolution by showing that the oracle complexity for target location can be made optimal, by increasing the number of calls to the walk operator that incorporates the graph structure by a logarithmic factor.
Our algorithm does not require amplitude amplification.
An important ingredient of our algorithm is the implementation of multi-step quantum walks by graph powering, using a coin space of walk-length dependent dimension, which may be of independent interest.
Finally, we demonstrate how to implement quantum walks arising from powers of symmetric Markov chains using our methods.}{}{}

\vspace*{10pt}

\vspace*{3pt}
\communicate{R. Cleve \& A. Harrow}

\vspace*{1pt}\textlineskip    

\setcounter{footnote}{0}
\renewcommand{\thefootnote}{\alph{footnote}}


\section{Introduction}

Quantum spatial search is an extension of Grover search, with the items arranged in a spatially distributed database and locality constraints on how they may be explored.
Most algorithms for its solution work by alternatively applying a quantum walk operator and an oracle operator on a starting state, till the final state has a large overlap with a target state.
Then the final state is measured, and the location of a marked item is revealed with high probability.
Quantum search algorithms with locality constraints have been extensively used in the quantum information literature to design many query efficient algorithms \cite{ambainis2003rev}, and also as tools to prove interesting complexity theoretic results such as exponential quantum speed-up in an oracular setting \cite{childs2003walk} and fast gap amplification in QMA \cite{nagaj2009amp}.

The performance of quantum spatial search is constrained by the connectivity of the underlying database.
The optimal query complexity value of $O(\sqrt{N})$ is not achieved for all graphs.
For instance, the best known query complexity for quantum spatial search on a 2D grid is $O(\sqrt{N \log N})$, which was achieved by Tulsi \cite{tulsi2008faster}, building on earlier work by Ambainis, Kempe and Rivosh (AKR) \cite{ambainis2005coins}.
For more general graphs with multiple marked items, indirect variants of quantum search, which combine the phase estimation procedure with quantum walks, have good query complexity \cite{krovi2016walk,magniez2011walk}.
More recently, we have analyzed direct quantum search for regular graphs with multiple targets, and have found that the direct approach matches the performance of the indirect search algorithms \cite{abhijith2018spatial}.
For all these cases, the speed-up is quadratic or sub-quadratic over the corresponding classical random walk search.

Our main result in this work is a strategy to optimize the query complexity of spatial search on the 2D grid, at the cost of increasing the number of quantum walk steps by a logarithmic factor.
The walk operator can be implemented by an oracle that knows the underlying graph structure, and making multiple queries to this oracle corresponds to taking multiple steps of the walk.
Increasing the number of walk steps effectively increases the connectivity of the graph.
With a sufficient number of walk steps, the effective graph would become an expander having connectivity properties close to that of a complete graph, and then the query complexity would become $O(\sqrt{N})$.
Our results demonstrate that optimal query complexity can be achieved much before this expander graph stage, with $O(\sqrt{N}\log N)$ total quantum walk steps.
As we show, this happens because the complexity of quantum search depends on the whole spectrum of the graph and not just on its spectral gap.

Our algorithm introduces a form of multi-stepping in the coin space of the walk, which leads to a new quantum walk whose spectral properties are simply related to the spectral properties of the original quantum walk.
Multi-stepping in the context of quantum walks refers to applying the walk more than once between consecutive oracle operations.
Search algorithms that use this technique are usually difficult to analyze except for very specific cases.%
\footnote{For instance see the walk used by Ambainis in Ref.~\cite{ambainis2007quantum}, and also the results in Ref.~\cite{wong2015search}.}
~This is because the performance of the quantum search depends crucially on the spectral properties of the walk operator, and these properties become obscured by repeated powering of the walk operator.
Nevertheless, the properties of such multi-step walks have been studied numerically \cite{patel2010search1}, and improvements in performance of the algorithm have been observed.
We introduce a different kind of multi-stepping here compared to previous approaches in the literature.
It results in a graph on which two-dimensional quantum search is optimal despite having small algebraic connectivity, similar to the graphs studied in Ref. \cite{meyer2015walk} in a continuous time setting.

The result obtained by Magniez, Nayak, Roland and Santha (MNRS) \cite{magniez2011walk} is pertinent to our work, as it also optimizes the query complexity by taking multiple quantum walk steps.
The MNRS algorithm is an indirect search algorithm, and is quite different from our proposal.
We will compare the performance of our algorithm with the MNRS algorithm while discussing quantum search on general graphs.

This article is organized as follows.
In Section \ref{sec:twoquery}, we explain how multi-stepping leads to two types of query complexity.
Our notation and the graph-powering method are presented in Section \ref{sec:prelim}.
We obtain the spectrum of the multi-step quantum walk in Section \ref{sec:walkspectrum}.
Our analysis of quantum search on a two-dimensional grid is presented in \ref{sec:search2d}.
Section \ref{sec:markovchain} describes how our multi-step walk technique can be applied to general Markov chains, while two appendices contain technical proofs of the bounds used in our analysis.

\section{Two query complexities}
\label{sec:twoquery}

The quantum spatial search algorithms of AKR \cite{ambainis2005coins} and Tulsi \cite{tulsi2008faster} work by alternatively applying a local walk operator ($W$) and an oracle operator ($O$) that marks the target state.
A particular choice of the quantum walk operator is the flip-flop walk \cite{shenvi2003quantum}.
It is a product of two reflection operators, the shift operator ($S$) and a coin operator ($C$).
The full search operator that is iterated in the algorithm is then,
\begin{equation}
U = WO = SCO ~.
\end{equation}

The number of iterations of $U$ required to find the target state is the query complexity of the algorithm, and equals the number of times the oracle encoded in $O$ is called.
But with closer scrutiny, we can define another oracle buried inside the shift operator, which provides the connectivity information of the graph when invoked.
Thus one can have separate query complexities corresponding to both these oracles.
In the AKR and Tulsi's algorithms, both these oracles are invoked only once per iteration, and hence their query complexities match.

We denote the query complexity for the target state oracle, $O$, as $Q_O$, and the query complexity for the graph structure oracle as $Q_G$.
For spatial search on a two-dimensional grid, the AKR algorithm has $Q_O = Q_G = \Theta(\sqrt{N}\log N)$, and Tulsi's algorithm improves that to $Q_O = Q_G = \Theta(\sqrt{N\log N})$.
The optimal value of $Q_O = \Theta(\sqrt{N})$ is achieved by Grover's algorithm, which corresponds to search on a complete graph without any locality constraints.
It is an open problem whether spatial search on a two-dimensional grid can achieve this optimal value.
In this work, we show that $Q_O$ can be made optimal at the expense of increasing $Q_G$ by logarithmic factors.

The central idea of our algorithm is to increase the connectivity of the graph by the technique of graph powering.
It corresponds to taking multiple walk steps on the graph before applying the target state oracle, i.e. $U_t = W^t O$.
Increasing the graph connectivity by appropriately increasing $t$, we can make $Q_O$ attain its optimal value.
One may guess that the required value of $t$ would scale as the the diameter of the graph, but we show that it is only logarithmic in the size of the graph. 

A multi-step classical random walk can be easily implemented by taking higher powers of the random walk operator.
But this procedure is not effective in case of quantum walks, because the walk operator is unitary and its eigenvalues show periodic behaviour when powered.
As we show in this work, this difficulty can be circumvented by using a coin space whose dimension increases with the number of walk steps.
The eigenvalues of this modified walk operator show a much more controlled behaviour when the underlying graph is powered, in contrast to quantum walk operators that use a coin space of fixed dimension.

\section{Preliminaries and notation}
\label{sec:prelim}

\subsection{Graph properties}

We consider a $d$-regular, undirected graph $G(V,E)$.
Here $V$ is the vertex set and $E$ is the edge set of the graph.
We denote the size of $V$ by $N$, i.e. $|V|=N$, while the size of $E$ is $|E|=dN/2$.
The adjacency matrix $A_G$ of this graph is an $N \times N$ matrix that encodes the connectivity information of the graph.
We use the normalized adjacency matrix for our work:
\begin{equation}
(A_G)_{ij} = \begin{cases}
         \frac{1}{d} ~,~~ \text{if}~ i,j \in E,\\
         0 ~,~~ \text{otherwise}.
         \end{cases}		
\end{equation}
$A_G$ is a real symmetric matrix.
So all its eigenvalues are real, all its eigenvectors can be chosen to be real, and they form a complete orthonormal set.
With our normalization, the eigenvalues of $A_G$ lie in the interval $[-1,1]$ and hence can be expressed as $\cos(\phi)$ for some $\phi\in[0,\pi]$.
We call the angle $\phi$ an eigenphase of $A_G$.
This adjacency matrix is related to the discrete Laplacian for the graph,
\begin{equation}
\Delta_G = A_G - \mathbb{I} ~.
\end{equation}
The largest eigenvalue of $A_G$ is always $1$, and the corresponding eigenvector is the uniform superposition vector.

$A_G$ is a doubly stochastic matrix, as every row and column of $A_G$ adds up to $1$:
\begin{equation}
\sum_i (A_G)_{ij} =  \sum_j (A_G)_{ij} = 1.
\end{equation}
It follows that any positive power of $A_G$ is also doubly stochastic, in addition to being a real symmetric matrix.
We use positive powers of $A_G$ throughout this work, and denote the set of all $t$ step paths in $G$ by $E^t$.
The value of $(A^t_G)_{ij}$ is the number of $t$ step paths connecting $i$ and $j$, divided by $d^t$.

\subsection{Shift operator and rotation maps}

The connectivity information of the graph is provided by the shift operator $S$, acting in a space of $dN$ dimensions.
For the two-dimensional grid, with vertices labeled as $(x,y)$ and four directions at each vertex, $S$ is defined by:
\begin{eqnarray}
S\ket{x,y,\rightarrow} = \ket{x+1,y,\leftarrow},~~ S\ket{x,y,\leftarrow} = \ket{x-1,y,\rightarrow}, \\
S\ket{x,y,\uparrow} = \ket{x,y+1,\downarrow},~~ S\ket{x,y,\downarrow} = \ket{x,y-1,\uparrow}.
\end{eqnarray}
It is possible to define shift operators for any regular graph \cite{abhijith2018spatial}.

In the computer science literature, a function that encodes the locality structure of the graph is called a rotation map $\mathcal{R}_G$ \cite{reingold2000entropy}.
For an $N$ vertex graph that is $d$ regular, the rotation map is a function $\mathcal{R}_G: [N]\times[d] \rightarrow [N]\times[d]$.
When $v$ is a neighbour of $u$, and the edge that connects them is labeled by $g$ at $u$ and $h$ at $v$, then $\mathcal{R}_G (u,g) = (v,h)$.
From the definition of $S$, it is clear that it encodes such a rotation map on the grid, albeit with a more natural labeling of the edges.

\subsection{Graph powering} 

The shift operator $S$ acts in a space of dimension $dN$, and encodes the structure of the graph $G$ via $\mathcal{R}_G$.
Now we define a new shift operator on an expanded space of size $d^t N$, which encodes the structure of the graph $G^t$, obtained from $G$ by taking its $t^{\text{th}}$ power.

The powering of a graph is done in terms of the normalized adjacency matrix $A_{G}$, by viewing $A^t_{G}$ as the normalized adjacency matrix for a $d^t$-regular graph.
This new graph is labeled $G^t$.
The vertices in $G^t$ are the same as the vertices in $G$, but the edges in $G^t$ correspond to paths of length $t$ in $G$.
Note that $G^t$ can have multiple edges between the same vertices, and also self-loops, even if $G$ doesn't have these.
Also by definition, $A_{G^t} =  (A_{G})^t$. 

The rotation map of the powered graph, $\mathcal{R}_{G^t}$, can be computed by $t$ applications of $\mathcal{R}_G$.
It lifts the degeneracy between paths with the same end points.
Specifically, $\mathcal{R}_{G^t}: [N]\times[d]^t \rightarrow [N]\times[d]^t$, such that if $\mathcal{R}_{G^t} (u,g_1,\ldots,g_t) = (v,h_1,\ldots,h_t)$, then $u$ and $v$ are connected by a path of length $t$ in $G$, and the edges of this path are labeled $g_1$ to $g_t$ when going from $u$ to $v$ and $h_1$ to $h_t$ when going from $v$ to $u$.
An illustration is provided in \figurename{~\ref{fig:1}}.

\begin{figure}[th]
{
\begin{center}

\begin{picture}(120,100)
\put(20,10){\circle*{4}}
\put(20,50){\circle*{4}}
\put(60,50){\circle*{4}}
\put(60,90){\circle*{4}}
\put(100,90){\circle*{4}}

\put(20,10){\line(0,1){40}}
\put(20,50){\line(1,0){40}}
\put(60,50){\line(0,1){40}}
\put(60,90){\line(1,0){40}}

\put(20,10){\vector(0,1){10}}
\put(20,50){\vector(1,0){10}}
\put(60,50){\vector(0,1){10}}
\put(60,90){\vector(1,0){10}}

\put(20,50){\vector(0,-1){10}}
\put(60,50){\vector(-1,0){10}}
\put(60,90){\vector(0,-1){10}}
\put(100,90){\vector(-1,0){10}}

\put(25,8){$u$}
\put(105,88){$v$}
\put(6,16){$g_1$}
\put(6,40){$h_4$}
\put(26,56){$g_2$}
\put(48,40){$h_3$}
\put(66,56){$g_3$}
\put(46,82){$h_2$}
\put(66,96){$g_4$}
\put(90,96){$h_1$}
\end{picture}
\end{center}
}
\fcaption{\label{fig:1}An edge in $G^4$ such that $\mathcal{R}_{G^4} (u, g_1, g_2, g_3, g_4) = (v, h_1, h_2, h_3, h_4)$.}
\end{figure}

We can now define the shift operator $S_t$, using $\mathcal{R}_{G^t}$:
\begin{equation}
S_t \ket{u,g_1,\ldots,g_t} = \ket{v,h_1,\ldots,h_t}, ~~\text{if}~~ \mathcal{R}_{G^t} (u,g_1,\ldots,g_t) = (v,h_1,\ldots,h_t).
\end{equation}
$S_t$ is a reflection operator and can be implemented using $t$ uses of $S$, as follows.
First apply $S$ to $u$ and $g_1$.
That reverses the direction of $g_1$, and shifts $u$ to the second vertex in the path.
Next apply $S$ to the vertex in the first register and $g_2$, and so on till we reach $g_t$.
At this point, the vertex in the first register will be $v$, but the registers with the edge labels will be in the reverse direction.
So to obtain $S_t$, we reverse the order of all the edge labels.
Thus a single application of $S_t$ requires $t$ calls to the function $\mathcal{R}_G$.

We construct our quantum spatial search algorithm in the Hilbert space $\mathcal{C}^N \otimes \mathcal{C}^{d^t}$, associated with the powered graph $G^t$.
In this Hilbert space, we define the coin operator,
\begin{equation}
C_t = 2\sum_{u}\ket{\psi^t_u}\bra{\psi^t_u} - \mathbb{I},
\end{equation}
where $\ket{\psi^t_u} = \frac{1}{d^{t/2}}\sum_{\bf{g}\in[d]^t} \ket{u,\bf{g}}$.
It is a simple extension of the definition of the coin operator on the graph $G$, i.e. $C = 2\sum_{u}\ket{\psi_u}\bra{\psi_u} - \mathbb{I}$ with $\ket{\psi_u} = \frac{1}{d^{1/2}}\sum_{g\in[d]^t} \ket{u,g}$.
$C_t$ is also a reflection operator, and can be implemented by applying $C$ to each of the $t$ edge labels.

Our quantum walk operator on the graph $G^t$ is $W_t = S_t C_t$.
A search algorithm that uses such a walk operator between oracle calls has different values of $Q_O$ and $Q_G$.
Since we need $t$ calls to $\mathcal{R}_G$ for each step of $W_t$, we have $Q_G = t Q_O$.

\section{Spectrum of $W_t$}
\label{sec:walkspectrum}

To understand the properties of $W_t$, we define a complete orthonormal basis for the Hilbert space $\mathbb{C}^N \otimes \mathbb{C}^{d^t}$, which captures the connectivity structure of $G^t$.
Let ${\bf p}$ be a $t$ step path in $G$ connecting vertices $u$ and $v$; alternatively it is an edge in $G^t$, and we label it as an element of $E^t$.
Let the sequence of edge labels along this path be $g_1,\ldots,g_t$ as we go from $u$ to $v$, and $h_1,\ldots,h_t$ as we go from $v$ to $u$.
Then we define a complete orthonormal basis for $\mathbb{C}^{Nd^t}$, with two basis vectors corresponding to each such path, as
\begin{equation}
\ket{{\bf p}^\pm} = \frac{1}{\sqrt{2}} ( \ket{u, g_1, \ldots g_t} \pm  \ket{v, h_1, \ldots, h_t}).
\end{equation}
All these vectors are the eigenvectors of the shift operator $S_t$, with $S_t \ket{{\bf p}^\pm} = \pm\ket{{\bf p}^\pm}$.%
\footnote{Some of the $\ket{{\bf p}^-}$ vectors may vanish, when the contributing paths double back on themselves.
Such a situation can be avoided by choosing $t$ to be odd.}

Now we prove an important theorem that helps us completely characterize the eigenvectors and eigenvalues of $W_t$ in terms of those of $A_{G^t}$.

\smallskip
\begin{theorem}\label{theo_1}
Let $W_t$ satisfy the eigenvalue equation $W_t\ket{\Phi^t_k} = e^{i\phi^{(t)}_k}\ket{\Phi^t_k}$, and let $a_{ku} = \braket{\Phi^t_k|\psi^t_u}$.
Then the vector $\vec{a}_k = (a_{k1},a_{k2},\ldots,a_{kN})$ is an eigenvector of $A_{G^t}$, with eigenvalue $cos(\phi^{(t)}_k)$, when $0<\phi^{(t)}_k<\pi$ (i.e. for all eigenvectors of $W_t$ which do not have eigenvalues $\pm1$). 

(\textbf{Converse}) Also, for every eigenvector of $A_{G^t}$ with eigenvalue $\cos(\phi^{(t)})$ and $0<\phi^{(t)}<\pi$, there exist two eigenvectors of $W_t$ with eigenvalues $e^{i\phi^{(t)}}$ and $e^{-i\phi^{(t)}}$. 
\end{theorem}

\noindent{\bf Proof:}
The proof closely follows the logic of Ref.~\cite{abhijith2018spatial}, where similar relations have been found between properties of classical and quantum walks.
The key idea is to find expressions for the components of $\ket{\Phi_k^t}$ in the $\ket{{\bf p}^\pm}$ basis that implicitly captures the locality structure of the graph. 
The same technique works here as $G^t$ is a regular graph with paths of length $t$ acting as edges.
The main difference from the analysis in Ref.~\cite{abhijith2018spatial} is that $G^t$ can contain multiple edges between the same vertices, but this is not a hurdle once we take into account that the number of paths of length $t$ connecting vertices $i$ and $j$ is $d^t$ times the matrix element $(A^t_G)_{ij}$.
We outline the proof below to set up the notation, while skipping the details.

With $W_t = S_t C_t$, we compute the components of the eigenvalue equation in the path basis by evaluating the matrix elements $\braket{\textbf{p}^\pm|\Phi^t_k}$.
\begin{equation}
e^{i\phi^{(t)}_k} \braket{\textbf{p}^+|\Phi^t_k}
  = \braket{\textbf{p}^+| S_t C_t |\Phi^t_k} 
  = \braket{\textbf{p}^+|2\sum_{u}\ket{\psi^t_{u}}\bra{\psi^t_{u}} - \mathbb{I}|\Phi^t_k} .
\end{equation}
These can be rewritten as,
\begin{equation}
(e^{i\phi^{(t)}_k}+1) \braket{\textbf{p}^+|{\Phi}^t_k}
  = 2 \sum_u \braket{\textbf{p}^+|\psi^t_u} a_{ku}^* .
\label{eq:comp_constraint}
\end{equation}
For a path $\textbf{p}$ that has vertices $u$ and $v$ as the end points, we have the result,
\begin{equation}\label{eq:component}
\braket{\Phi^t_k|\textbf{p}^+} = \sqrt{\frac{2}{d^t}}~ \frac{a_{ku}+a_{kv}}{1+e^{-i\phi^{(t)}_k}} ~.
\end{equation}
Similarly,
\begin{equation}
\braket{\Phi^t_k|\textbf{p}^-} = \sqrt{\frac{2}{d^t}}~ \frac{a_{ku}-a_{kv}}{1-e^{-i\phi^{(t)}_k}} ~.
\end{equation}
There is a sign ambiguity in this expression arising from the ordering of $u$ and $v$ in the definition of $\ket{\textbf{p}^-}$, but it disappears from the final results.

Since $\ket{\textbf{p}^\pm}$ form a complete basis, $\ket{\Phi^t_k}$ is completely determined in terms of the values of $a_{ku}$ and $\phi^{(t)}_k$.
So, inserting the values of the overlaps, we get
\begin{align}
a_{ki} &= \sum_{{\bf p}\in E^t}  \braket{{\Phi}^{t}_k | \textbf{p}^+} \braket{\textbf{p}^+| \psi^t_i} +  \braket{{\Phi}^{t}_k | \textbf{p}^-} \braket{\textbf{p}^-| \psi^t_i} ~,\\
       & = \sum_{j \in V} (A^t_G)_{ij} (\frac{a_{ki} + a_{kj}}{1+e^{-i\phi^{(t)}_k}} + \frac{a_{ki}-a_{kj}}{1-e^{-i\phi^{(t)}_k}}) ~,\\
       & = a_{ki} \left(\frac{1}{1+e^{-i\phi^{(t)}_k}} + \frac{1}{1-e^{-i\phi^{(t)}_k}}\right) + \sum_{j \in V} (A^t_G)_{ij} a_{kj} \left(\frac{1}{1+e^{-i\phi^{(t)}_k}}-\frac{1}{1-e^{-i\phi^{(t)}_k}}\right) ~.
\label{eq:aki_one}
\end{align} 
In the last step, we have used the fact that $A_G^t$ is a doubly stochastic matrix.

We can derive another expression for $a_{ki}$, using the fact that $\ket{\psi^t_i}$ is an eigenvector of $C_t$ with eigenvalue $1$.
Unitarity of $W_t$ implies $W_t^\dagger S_t = W_t^{-1}S_t = C_t^{-1}S_t^{-1}S_t = C_t$.
So,
\begin{align}
a_{ki} &= \braket{\Phi^t_k |C_t| \psi^t_i} ~,\\
       &= \sum_{{\bf p} \in {E}^t } {\Big (}  e^{-i {\phi}^{(t)}_k} \braket{{\Phi}^{t}_k | \textbf{p}^+} \braket{\textbf{p}^+| \psi^t_i} - e^{-i {\phi}^{(t)}_k}  \braket{{\Phi}^{t}_k | \textbf{p}^-} \braket{\textbf{p}^-| \psi^t_i}{\Big )} .
\end{align}
Repetition of the steps leading to Eq.\eqref{eq:aki_one} then gives,
\begin{align}
a_{ki} &= a_{ki}~e^{-i\phi^{(t)}_k} \left(\frac{1}{1+e^{-i\phi^{(t)}_k}}-\frac{1}{1-e^{-i\phi^{(t)}_k}}\right) \\
       &+ \sum_{j \in V} (A^t_G)_{ij}~a_{kj}~e^{-i\phi^{(t)}_k} \left(\frac{1}{1+e^{-i\phi^{(t)}_k}}+\frac{1}{1-e^{-i\phi^{(t)}_k}}\right) ~.
\label{eq:aki_two}
\end{align}

Subtraction of Eq.\eqref{eq:aki_two} from Eq.\eqref{eq:aki_one}, and then rearrangement of the terms, yield the eigenvalue equation,
\begin{equation}
\sum_{j \in V} (A^t_G)_{ij} ~a_{kj} =  \cos({\phi}^{(t)}_k) ~ a_{ki} ~.
\end{equation}

\paragraph{Proof of converse:}
Let $\vec{a}$ be an eigenvector of $A_{G^t}$ with eigenvalue $\cos(\phi^{(t)})$.
Let $a_u$ be the component of $\vec{a}$ at the vertex $u$.
Now we define a vector $\ket{\Phi} \in C^N \otimes C^{d^t}$ such that its components in the path basis are,
\begin{equation}
\braket{{\Phi}|\textbf{p}^\pm} = \sqrt{\frac{2}{d^t}} ~\frac{a_u\pm a_v}{1\pm e^{-i\phi^{(t)}}} ~,
\end{equation}
where $\textbf{p}$ connects $u$ and $v$.

The uniform superposition state of all the paths starting at a vertex $u$ can be written as,
\begin{equation}
\ket{\psi^t_u} = \frac{1}{d^{t/2}} \sum_{{\bf g}\in[d]^t} \ket{u,{\bf g}}
             = \frac{1}{\sqrt{2d^t}} \sum_{\textbf{p}=(u,w)} (\ket{\textbf{p}^+}+\ket{\textbf{p}^-}) ~.
\end{equation}
Therefore,
\begin{align}
\braket{\psi^t_u|\Phi} &= \sum_{w \in V } ({A^t_G})_{uw} \left( \frac{a_u+a_w}{1+e^{i\phi^{(t)}}} + \frac{a_u-a_w}{1-e^{i\phi^{(t)}}} \right) ~,\\
 &= a_u \left( \frac{1}{1+e^{i\phi^{(t)}}} + \frac{1}{1-e^{i\phi^{(t)}}} \right) + \cos(\phi^{(t)})~a_u \left( \frac{1}{1+e^{i\phi^{(t)}}} - \frac{1}{1-e^{i\phi^{(t)}}} \right) ~,
\end{align}
where in the last step we have used the eigenvalue equation, $\sum_w (A^t_G)_{uw}a_w = \cos(\phi^{(t)})~a_u$.
Simplifying further, we find $\braket{\psi^t_u|\Phi} =  a_u$.

Now we are ready to look at the action of $W_t$ on $\ket{\Phi}$ in the path basis.
Let $\textbf{p}$ be a path connecting $u$ and $v$.
Then
\begin{align}
\braket{\textbf{p}^\pm|W_t|\Phi} = \pm\braket{\textbf{p}^\pm|C_t|\Phi}
  &= \pm\sqrt{\frac{2}{d^t}} \left( \braket{\psi^t_u|\Phi} \pm \braket{\psi^t_v|\Phi} \right) \mp \braket{\textbf{p}^\pm|\Phi} ~,\\
  &= e^{i\phi^{(t)}} \sqrt{\frac{2}{d^t}}~\frac{a_u \pm a_v}{1 \pm e^{i\phi^{(t)}}} ~,\\
  &= e^{i\phi^{(t)}} \braket{\textbf{p}^\pm|\Phi} ~.
\end{align}
So it is clear that $\ket{\Phi}$ is an eigenvector of $W_t$ for with eigenvalue  $e^{i\phi^{(t)}}$.
By taking the complex conjugate of $\ket{\Phi}$, we get the eigenvector with eigenvalue $e^{-i\phi^{(t)}}$.
This proves the converse of our theorem.
\hfill $\square\,$

\smallskip
This theorem has two important corollaries which make it possible to improve the query complexity of quantum search in 2D.
The first one is a trivial consequence of the theorem and the fact that $A_{G^t}=(A_G)^t$.

\begin{corollary}\label{cor:1}
$\cos({\phi}^{(t)}_k) = \cos^t(\phi_k)$, for $\phi_k\in(0,\pi)$.
\end{corollary}

We point out that the appearance of $\cos^t(\phi_k)$ is a key consequence of graph powering, which can be bounded easily.
Powering just the walk operator to $W^t$ produces $\cos(t\phi_k)$, which is not easy to bound.

The second, somewhat non-trivial corollary is that the coefficients $a_{ki}$ are independent of $t$.
Since the eigenvectors of $A_{G^t}$ are the same as that of $A_G$, one would expect $a_{ki}$ to be independent of $t$ up to an overall proportionality factor.
But we show that even this factor is independent of $t$.
We state this important result as a corollary.

\begin{corollary}\label{cor:2}
For $0<\phi^{(t)}_k<\pi$, and for all odd integers $t$,
\begin{equation}
\sum_u |\braket{\Phi^{t}_k | \psi^t_u}|^2 =  \sum_u a^2_{ku} = \frac{1}{2} ~.
\end{equation}
\end{corollary}

\noindent{\bf Proof:}
We impose the normalization condition on the eigenvectors $\ket{\Phi^t_k}$,
\begin{align}
1 &= \|\ket{\Phi^t_k}\|^2 ,\\
  &= \sum_{{\bf p}\in E^t} |\braket{\Phi^t_k|\textbf{p}^+}|^2 + |\braket{\Phi^t_k|\textbf{p}^-}|^2, \\
  &= \sum_{(u,v)\in V\times V} \left( (A^t_G)_{uv} ~\left| \frac{a_{ku}+a_{kv}}{1+e^{-i\phi^{(t)}_k}} \right|^2 + (A^t_G)_{uv} ~\left|\frac{a_{ku}-a_{kv}}{1-e^{-i\phi^{(t)}_k}} \right|^2 \right) ~.
\end{align} 
Here we have used our earlier expression for the components of $\ket{\Phi_k}$ in the $\ket{{\bf p}^\pm}$ basis, and also the fact that the number of length $t$ paths between two vertices are proportional to the corresponding element of $A^t_G$.
An extra factor of $2$ is taken care of by the double summation over $u,v$.
Since $a_{ki}$ are the components of an eigenvector of a real symmetric matrix, we have chosen them to be real.
Thus,
\begin{equation}
4 = \sum_{u,v} \left( (A^t_G)_{uv} ~\frac{(a_{ku}+a_{kv})^2}{\cos^2(\phi^{(t)}_k/2)} + (A^t_G)_{uv} ~\frac{(a_{ku}-a_{kv})^2}{\sin^2(\phi^{(t)}_k/2)} \right) ~.
\end{equation}   
Rearranging the terms, we get the result,
\begin{equation}
\sum_{u,v} \left( (A^t_G)_{uv}~a^2_{ku} + (A^t_G)_{uv}~a^2_{kv} - 2a_{ku}~(A^t_G)_{uv}~a_{kv}\cos(\phi^{(t)}_k) \right) = \sin^2(\phi^{(t)}_k) ~.
\end{equation}
We know that $\sum_{v} (A^t_G)_{uv} = \sum_{u} (A^t_G)_{uv} = 1$, and $\sum_{u,v} a_{ku}(A^t_G)_{uv}a_{kv} = \cos(\phi^{(t)}_k) \sum_u a_{ku}^2$.
Using these conditions, we obtain the projection condition,
\begin{equation}\label{eq:normalize}
\sum_u a^2_{ku} = \frac{1}{2} ~.
\end{equation}
This condition is independent of $t$, and we choose $t$ to be odd in order to avoid zero length paths.
\hfill $\square\,$

Thus the values of $a_{ku}$ are completely independent of $t$.
The projection is less than one because the states $\ket{\psi^t_u}$ do not span $\mathbb{C}^N \otimes \mathbb{C}^{d^t}$.
What we have shown is that the magnitude of the projections of the eigenvectors of $W_t$, on to the subspace spanned by $\{\ket{\psi^t_u} ~|~ u\in V\}$, is a constant. 

Now we proceed to analyze the performance of $W_t$ for search on a two-dimensional grid with periodic boundary conditions.

\section{Quantum search on a two-dimensional grid with $W_t$}
\label{sec:search2d}

The quantum search algorithm requires an oracle that marks the target states.
Following the perturbed coin approach of AKR \cite{ambainis2005coins}, we define the oracle $O_t$ as,
\begin{equation}\label{eq:oracle_def}
O_t = \mathbb{I} - 2 \ket{\psi^t_m}\bra{\psi^t_m} ~.
\end{equation}
Here $m \in V$ labels the marked states that we want to locate.
The abstract quantum search algorithm then proceeds as follows.
The search operator $U_t = W_t O_t$ is applied to the starting state, $\ket{\Psi_S}$, $Q$ times.
The resultant state, up on measurement, collapses to the target state with some success probability.
If this success probability is not large enough, one possibility is to boost it to a constant value, with sufficient rounds of amplitude amplification.
A better alternative is to use Tulsi's controlled search technique \cite{tulsi2008faster}, which enhances the success probability to a constant without any need for amplitude amplification.
The algorithm is always optimized by minimizing the total number of required steps.
 
\subsection{Abstract search framework}

We take the starting state of our algorithm to be the uniform superposition state, which also happens to be an eigenvector of $W_t$ with eigenvalue $1$,
\begin{equation}\label{eq:starting_state}
\ket{\Psi_S} =  \ket{\Phi_0^t} =  \frac{1}{\sqrt{N}} \sum_{u \in V} \ket{\psi^t_u} ~.
\end{equation}
Note that for $t=1$, our algorithm reduces to the flip-flop quantum search.
To analyze it for general $t$, we first show that quantum search using $U_t$ fits into the abstract search framework constructed by AKR \cite{ambainis2005coins}.
This framework requires a subspace $\mathcal{H} \in \mathbb{C}^{Nd^t}$ such that,\\
$\bullet$ $\mathcal{H}$ is invariant under $U_t$.\\
$\bullet$ $W_t$ is expressible as a real operator on $\mathcal{H}$.\\
$\bullet$ Only one eigenvector of $W_t$ corresponding to eigenvalue $1$ lies in $\mathcal{H}$, and it is the starting state $\ket{\Psi_S}$.

We take $\mathcal{H}$ to be the space spanned by $\ket{\Phi^t_0}$ and all the eigenvectors of $W_t$ with non-real eigenvalues.
As a consequence of Theorem \ref{theo_1}, this subspace has dimension $2N-1$.%
\footnote{This is true only when $G_t$ is not bipartite.
For bipartite graphs, one of the complex eigenvalue pairs of $W_t$ is replaced by a $-1$ eigenvalue, and the subspace has dimension $2N-2$.
The eigenvector $|\Phi_b^t\rangle$ corresponding to the $-1$ eigenvalue has the same structure as $|\Phi_0^t\rangle$, except that the coefficients for the two partitions have opposite sign.
This case is analyzed in detail in Ref.~\cite{abhijith2018spatial}.}
~Then the last two conditions listed above are satisfied in $\mathcal{H}$ by construction.
Moreover, since $\mathcal{H}$ is invariant under $W_t$, the first condition is satisfied provided $\mathcal{H}$ is invariant under $O_t$ as well.
This is indeed the case, because $O_t$ is a reflection about the state $\ket{\psi^t_m}$, and we show in the following lemma that all the states of the form $\ket{\psi^t_u}$ lie in $\mathcal{H}$.

\smallskip
\begin{lemma}
For all $u$ and odd integers $t$, $\ket{\psi^t_u} \in \mathcal{H}$.
\end{lemma}

\noindent{\bf Proof}:
The part of $\mathbb{C}^{Nd^t}$ orthogonal to $\mathcal{H}$ is spanned by eigenvectors of $W_t$ with eigenvalues $\pm1$.
Let $\ket{\Phi_l}$ to be an eigenvector of $W_t$ with eigenvalue $1$.
Then using the same steps that were used to derive Eq.\eqref{eq:comp_constraint}, we find
\begin{equation}
\sum_{u} a_{lu} \braket{\psi^t_u|\mathbf{p}^-} = 0 ~.
\end{equation} 
It implies that for any $u$ and $v$ connected by a length $t$ path, $a_{lu}-a_{lv}=0$.
Since the two-dimensional grid is strongly connected, and we have chosen $t$ to be odd,%
\footnote{Odd $t$ allows neighbouring vertices of $G$ to be connected by a length $t$ path.}
~this means that $a_{lu}$ must be the same for all $u$.
When the common value of $a_{lu}$ is nonzero, we have $\ket{\Phi_l}=\ket{\Phi^t_0}$.
Otherwise, $a_{lu}=0$ implies that all such eigenvectors of $W_t$ with eigenvalue $1$ are orthogonal to the states of the form $\ket{\psi^t_u}$.
 
Similarly, by choosing $\ket{\Phi_l}$ to be an eigenvector of $W_t$ with eigenvalue $-1$, and considering its overlap with $\ket{\mathbf{p}^+}$, we find that $a_{lu}+a_{lv}=0$ for any $u$ and $v$ connected by a length $t$ path.
For a bipartite graph, it is possible that the components $a_{lu}$ are nonzero with their values alternating in sign, but then we have $\ket{\Phi_l}=\ket{\Phi^t_b}$.
Otherwise, $a_{lu} = 0$ for all $u$, which implies that all such eigenvectors of $W_t$ with eigenvalue $-1$ have no overlap with the states of the form $\ket{\psi^t_u}$.
 
Thus all the states of the form $\ket{\psi^t_u}$ are orthogonal to all the states that lie outside $\mathcal{H}$, and so $O_t$ leaves $\mathcal{H}$ invariant.
\hfill $\square\,$

\smallskip
With $\mathcal{H}$ being invariant under $U_t$, we can now analyze our quantum search algorithm within the abstract search framework.
For convenience, we first recollect the main results obtained by AKR in this framework; their detailed explanations can be found in Ref.~\cite{ambainis2005coins}.

Let $U = WO$ be the search operator in a subspace $\mathcal{H}$, satisfying the three conditions we have listed.
Let $\ket{\Psi_S}$ and $\ket{\Psi_T}$ be the starting state and the target state for the search problem respectively.
We define $\ket{\Theta_k}$ to be the eigenvectors of $W$ with eigenphases $\theta_k$ (which are arranged in ascending order), and call the overlap of these eigenvectors with $\ket{\Psi_T}$ as $a_k$ (which are chosen to be real by convention).
As before, we choose $\ket{\Psi_S}=\ket{\Theta_0}$, which is the only eigenvector of $W$ in $\mathcal{H}$ with eigenvalue $1$.
The only relevant eigenvalues of $U$ in this framework are $e^{\pm i\alpha}$, where $\alpha$ is the smallest non-zero eigenphase of $U$; we label the corresponding eigenvectors $\ket{\pm\alpha}$.
Three important results were proven by AKR in this abstract search framework, which we state without proof.%
\footnote{Our phase conventions \cite{abhijith2018spatial} are different from those of AKR.}

\paragraph{Result 1:} The eigenphase $\alpha$ scales as,
\begin{equation}\label{eq:r1}
\alpha = \Theta\left( \frac{1}{\sqrt{\sum_{k\neq0}\frac{a^2_k}{a^2_0} \frac{1}{1-\cos\theta_k}}} \right).
\end{equation}

\paragraph{Result 2:} The starting state has a high overlap with the subspace spanned by $\ket{\pm\alpha}$.
Let $\ket{w_S} = \frac{1}{\sqrt{2}} (\ket{\alpha}+\ket{-\alpha})$.
Provided $\alpha<\frac{1}{2}\theta_1$, the overlap of this state with the starting state is,
\begin{equation}\label{eq:r2}
|\braket{w_S|\Psi_S}| =  1 - \Theta\left( \alpha^4\sum_{k\neq0} \frac{a^2_k}{a_0^2} \frac{1}{(1-\cos\theta_k)^2} \right).
\end{equation}

\paragraph{Result 3:} Action of $U$ for $Q=\lfloor\frac{\pi}{2\alpha}\rfloor$ iterations on $\ket{w_S}$ produces a state close to $\ket{w_T} = \frac{1}{\sqrt{2}} (\ket{\alpha}-\ket{-\alpha})$.
Provided $\alpha<\frac{1}{2}\theta_1$, the overlap of this state with the target state is,%
\footnote{In the AKR paper, this result is reported with $\cot^2(\theta_k/4)$, but a more careful analysis of their proof improves it to $\cot^2(\theta_k/2)$.}
\begin{equation}\label{eq:r3}
|\braket{w_T| \Psi_T}| = \text{min}\left( \Theta\left( \frac{1}{\sqrt{\sum_{k\neq0 } a^2_k\cot^2(\theta_k/2)}} \right),1 \right).
\end{equation} 

These results can be combined to find the success probability of an abstract search algorithm after $\lfloor\frac{\pi}{2\alpha}\rfloor$ iterations.
   
\subsection{Analysis of spatial search with $U_t$ on a two-dimensional grid}

We take the two-dimensional grid to be of size $\sqrt{N}\times\sqrt{N}$.
The adjacency matrix of the grid is diagonalized using a Fourier transform to obtain its eigenvalues and eigenvectors.
We label the eigenvalues by a tuple $\mathbf{k} = (k_x,k_y)$, where $k_x$ and $k_y$ take integral values from $0$ to $\sqrt{N}-1$.%
\footnote{We use modulo $\sqrt{N}$ labels, so that $-k\equiv\sqrt{N}-k$.}
~The eigenvalues are,%
\footnote{For simplicity, we refer to $\phi_k^{(1)}$ as just $\phi_k$ from now on.}
\begin{equation}\label{eq:classical_evals}
\cos\phi_\mathbf{k} = \frac{1}{2}(\cos(\frac{2\pi k_x}{\sqrt{N}})+\cos(\frac{2\pi k_y}{\sqrt{N}})) ~,
 \end{equation} 
and the corresponding eigenvectors have components $\frac{1}{\sqrt{N}} \exp(\frac{2\pi i\mathbf{k\cdot x}}{\sqrt{N}})$ at the point $\mathbf{x}=(x,y)$.

As per the definition of the oracle in Eq.\eqref{eq:oracle_def}, we are searching for the state $\ket{\Psi_T} = \ket{\psi_m^t}$.
According to Theorem \ref{theo_1}, the overlap of this state with the eigenvectors of $W_t$ in $\mathcal{H}$ is proportional to the corresponding component of the eigenvectors of $A_G$.
In the standard convention, the Fourier coefficients are complex, while we have chosen $a_{ku}$ to be real.
This is not a problem, because Eq.\eqref{eq:classical_evals} shows that the eigenvectors with $\mathbf{k}$ and $-\mathbf{k}$ have the same eigenvalue, and so the corresponding complex conjugate coefficients can be mixed to get real coefficients.%
\footnote{Equivalently, we can replace the projections $a_{ku}^2$ by $|a_{ku}|^2$ in our formulae.}
~Then using the result of Corollary \ref{cor:2}, together with translational invariance of $W_t$ and appropriate choice of global phases, we find the overlaps to be,
\begin{equation}
a_{\mathbf{k}}^2 = \braket{\Phi_{\mathbf{k}}|\Psi_T}^2 = \braket{\Phi_{\mathbf{k}}|\psi_m^t}^2 = \frac{1}{2N} ~, 
\end{equation}
for $\mathbf{k}\neq0$.
We also have $a_0=\frac{1}{\sqrt{N}}$ from the definition of the starting state in Eq.\eqref{eq:starting_state}.

Until now we have not selected a specific value for $t$.
To proceed further, we demand $t$ to be some function of $N$, which is tuned to get the desired scaling behaviour of $Q_O$ and $Q_G$.
The performance of the search algorithm is governed by the grid sums involving the eigenphases of the adjacency matrix that occur in Eqs.(\ref{eq:r1}-\ref{eq:r3}).
For $t=O(\log N)$, as proved in Appendix A, these sums scale as,
\begin{equation}\label{eq:ss1}
\sum_{\mathbf{k}\ne0}\frac{a^2_{\mathbf{k}}}{a^2_{0}}~\frac{1}{1-\cos(\phi_{\mathbf{k}}^{(t)})} = \frac{1}{2}\sum_{\mathbf{k}\ne0} \frac{1}{1-\cos^t\phi_{\mathbf{k}}} = \Theta(\frac{N\log N}{t}) ~,
\end{equation}
\begin{equation}\label{eq:ss2}
\sum_{\mathbf{k}\ne0}\frac{a^2_{\mathbf{k}}}{a^2_{0}}~\frac{1}{(1-\cos(\phi_{\mathbf{k}}^{(t)}))^2} = \frac{1}{2}\sum_{\mathbf{k}\ne0} \frac{1}{(1-\cos^t\phi_{\mathbf{k}})^2} = \Theta(\frac{N^2}{t^2}) ~,
\end{equation}
\begin{equation}\label{eq:ss3}
\sum_{\mathbf{k}\ne0}a^2_{\mathbf{k}} \cot^2(\phi_\mathbf{k}^{(t)}/2) = \frac{1}{2N}\sum_{\mathbf{k}\ne0} \cot^2(\phi_\mathbf{k}^{(t)}/2) = \Theta(\frac{\log N}{t}) ~.
\end{equation}
Using these expressions in Eqs.(\ref{eq:r1}-\ref{eq:r3}), we get under the condition that $t=O(\log N)$,
\begin{equation}\label{eq:s1}
\alpha = \Theta\left( \sqrt{\frac{t}{N\log N}} \right) ,
\end{equation}
\begin{equation}\label{eq:s2}
 |\braket{w_S|\Psi_S}| = 1-\Theta(\frac{1}{\log^2 N}) ~,
\end{equation}
\begin{equation}\label{eq:s3}
 |\braket{w_T|\Psi_T}| = \Theta\left( \sqrt{\frac{t}{\log N}} \right) .
\end{equation}
The last two equations are valid only when $\alpha<\frac{1}{2}\phi^{(t)}_1$.
We show in Appendix B that this condition holds asymptotically.

With these results, we can determine the performance of the quantum search algorithm that applies $U_t$ on $\ket{\Psi_S}$, $Q=\lfloor\frac{\pi}{2\alpha}\rfloor$ times.
There are three sources of error that limit the success of the algorithm: (i) $\frac{\pi}{2\alpha}$ may not be an integer, (ii) the initial state $|\Psi_S\rangle$ may have a component outside the $|\pm\alpha\rangle$ subspace, and (iii) the final state $(U_t)^Q|\Psi_S\rangle$ may not coincide with $|\Psi_T\rangle$.
Consequently, the success probability of the algorithm is lower bounded by a product of three factors,%
\footnote{After $Q$ iterations, the projection of the state in the subspace spanned by $\ket{\pm\alpha}$ is $\frac{1}{\sqrt{2}}(e^{iQ\alpha}\ket{\alpha}+e^{-iQ\alpha}\ket{-\alpha})$.
So truncation of $Q$ to $\lfloor\frac{\pi}{2\alpha}\rfloor$ can make the phase $Q\alpha$ differ by at most $\alpha$ from its optimal value $\frac{\pi}{2}$, which gives the first factor.
If $Q$ is rounded off to $[\frac{\pi}{2\alpha}]$ instead, the phase mismatch would be at most $\frac{\alpha}{2}$.}
\begin{equation}\label{eq:successprob}
p_s \ge \cos^2\alpha~|\braket{w_S|\Psi_S}|^2~|\braket{w_T|\Psi_T}|^2 = \Omega(\frac{t}{\log N}) ~,
\end{equation}
with the error being dominated by the third factor as per Eqs.(\ref{eq:s1}-\ref{eq:s3}).
 
To boost the success probability to a constant, we can use the amplitude amplification technique \cite{brassard2002quantum}.
The algorithm then requires $\Omega(\sqrt{\frac{\log N}{t}})$ rounds of amplification.
That makes the query complexity of the target state oracle, $Q_O =  O(\sqrt{\frac{\log N}{t}}~Q) = O(\frac{\sqrt{N}\log N}{t})$, and that for the graph structure oracle, $Q_G = tQ_O = O(\sqrt{N}\log N)$.  

In particular, by taking $t=\Theta(\log N)$, we have $p_s=\Theta(1)$, and the amplitude amplification step becomes unnecessary.
Then $Q_O$ achieves its optimal scaling behaviour $\Theta(\sqrt{N})$, while $Q_G$ remains independent of $t$.
Increasing $t$ beyond $\Theta(\log N)$ worsens $Q_G$, with no further improvement in $Q_O$.

\subsection{Multi-step version of Tulsi's algorithm}

The quantum search using $U_t$ can be seen as a multi-step version of the AKR algorithm.
A natural question to consider is whether its scaling can be further improved by Tulsi's technique of controlling the quantum walk and the oracle using an ancilla qubit \cite{tulsi2008faster}.
This technique can also be interpreted as adding a selective mass term at the marked vertex to a relativistic walk \cite{patel2010search2}, or as adding a self-loop to the graph at the marked vertex \cite{krovi2010search}.
We now show that we cannot further improve the performance of our quantum search algorithm using Tulsi's technique, but we can make $Q_G$ more controllable at the expense of $Q_O$.

\begin{figure}[th]
{
\begin{center}

\begin{picture}(180,90)
\put(20,60){\line(1,0){18}}
\put(54,60){\line(1,0){10}}
\put(68,60){\line(1,0){10}}
\put(94,60){\line(1,0){10}}
\put(108,60){\line(1,0){10}}
\put(134,60){\line(1,0){16}}
\put(20,20){\line(1,0){40}}
\put(72,20){\line(1,0){28}}
\put(112,20){\line(1,0){38}}

\put(66,60){\circle{4}}
\put(106,60){\circle{4}}
\put(66,26){\line(0,1){32}}
\put(106,26){\line(0,1){32}}

\put(38,52){\framebox(16,16){$X_\delta$}}
\put(78,52){\framebox(16,16){$X_\delta^\dagger$}}
\put(118,52){\framebox(16,16){$Z$}}
\put(60,14){\framebox(12,12){$O$}}
\put(100,14){\framebox(12,12){$W$}}

\put(0,18){$|\Phi_0^t\rangle$}
\put(4,57){$|0\rangle$}
\put(154,57){$|\delta\rangle$}
\put(154,18){$|\psi_m^t\rangle$}

\put(30,10){\dashbox{5}(110,65){}}
\put(90,80){\makebox(0,0)[b]{Iterate $Q_\delta$ times}}
\end{picture}
\end{center}
}
\fcaption{\label{fig:tulsi_circ}Quantum logic circuit for Tulsi's controlled spatial search algorithm.}
\end{figure}

Tulsi's controlled spatial search algorithm uses a tunable parameter $\delta$, and can be implemented using the quantum logic circuit shown in \figurename{~\ref{fig:tulsi_circ}}.
The starting state for the algorithm is $|\Phi_0^t\rangle|0\rangle$.
The single qubit operators are: 
\begin{equation}
X_\delta = \begin{pmatrix}
~\cos\delta & \sin\delta \\ 
-\sin\delta & \cos\delta
\end{pmatrix} ~,~~
Z = \begin{pmatrix} 1 & 0 \\ 0 & -1 \end{pmatrix} ~.
\end{equation}
After $Q_\delta$ iterations of the search operator, enclosed in the dashed box, the algorithm reaches a final state that has $\Theta(1)$ overlap with the target state $\ket{\psi_m^t}\ket{\delta}$, where $\ket{\delta}=X_\delta^\dagger\ket{0}$.

This modified algorithm can be analyzed in the abstract search framework with a new search operator $\tilde{U}_t=\tilde{W}_t\tilde{O}_t(\delta)$.
The new walk operator is,
\begin{equation}
\tilde{W}_t = \begin{pmatrix} W_t & 0\cr 0 & -\mathbb{I}\cr \end{pmatrix} ~,
\end{equation}
which adds an extra walk eigenstate with eigenvalue $-1$ to the invariant space $\mathcal{H}$, while the new oracle is,
\begin{equation}
\tilde{O}_t(\delta) = \mathbb{I} - 2\ket{\psi_m^t}\ket{\delta}\bra{\psi_m^t}\bra{\delta} ~,
\end{equation}
which searches for the state $\ket{\tilde{\Psi}_T(\delta)} = \ket{\Psi_T}\ket{\delta}$.

Let $a_\pi(\delta)$ be the overlap of the target state with the extra walk eigenstate with eigenvalue $-1$.
Inclusion of its contribution modifies the results in Eqs.(\ref{eq:r1}-\ref{eq:r3}) to \cite{tulsi2008faster}:
\begin{equation}\label{eq:t1}
\alpha_\delta = \Theta\left( \frac{a_0(\delta)}{\sqrt{\sum_{k\ne0}\frac{a^2_k(\delta)}{1-\cos\theta_k} + \frac{a^2_\pi(\delta)}{4}}} \right) ~,
\end{equation}
\begin{equation}\label{eq:t2}
|\braket{\tilde{w}_S|\tilde{\Psi}_S}| = 1 - \Theta\left( \alpha_\delta^4\sum_{k\ne0}\frac{a^2_k(\delta)}{a_0^2(\delta)}\frac{1}{(1-\cos\theta_k)^2} \right) - \Theta\left( \frac{\alpha_\delta^4 a^2_\pi(\delta)}{a^2_0(\delta)} \right) ~,
\end{equation}
\begin{equation}\label{eq:t3}
|\braket{\tilde{w}_T|\tilde{\Psi}_T}| = \text{min}\left(\Theta\left( \frac{1}{\sqrt{\sum_{k\ne0}a^2_k(\delta)\cot^2(\theta_k/2)}} \right),1 \right) ~.
\end{equation}
Note that the last equation remains the same as Eq.\eqref{eq:r3}.

The overlaps for our algorithm are,
\begin{equation}
a^2_{\mathbf{k}}(\delta) = |\braket{\tilde{\Psi}_T(\delta)|\tilde{\Phi}_{\mathbf{k}}}|^2 = a^2_\mathbf{k}\cos^2\delta ~,~~ a^2_\pi(\delta) = \sin^2\delta ~.
\end{equation}
So using the estimates in Eqs.(\ref{eq:ss1}-\ref{eq:ss3}), Eqs.(\ref{eq:t1}-\ref{eq:t3}) become:
\begin{equation}\label{eq:u1}
\alpha_\delta = \Theta\left( \frac{1}{\sqrt{\frac{N\log N}{t} + \frac{N}{4}\tan^2\delta}} \right) ~,
\end{equation}
\begin{equation}\label{eq:u2}
|\braket{\tilde{w}_S|\tilde{\Psi}_S}| =  1 - \Theta(\frac{\alpha_\delta^4N^2}{t^2}) - \Theta(\alpha_\delta^4N\tan^2\delta) ~,
\end{equation}
\begin{equation}\label{eq:u3}
|\braket{\tilde{w}_T|\tilde{\Psi}_T}| = \text{min}\left(\Theta\left( \sqrt{\frac{t(1+\tan^2\delta)}{\log N}} \right),1 \right) ~.
\end{equation}

We have assumed $t=O(\log N)$.
So Eq.\eqref{eq:u1} implies that $\alpha_\delta=O(\frac{1}{\sqrt{N}})$.
Then, following the analysis that led to Eq.\eqref{eq:successprob}, the quantum search algorithm with $Q_\delta=\lfloor\frac{\pi}{2\alpha_\delta}\rfloor$ iterations of $\tilde{U}_t$, succeeds with probability
\begin{equation}
p_s(\delta) = \Omega\left( \frac{t(1+\tan^2\delta)}{\log N} \right) ~.
\end{equation}
The choice $t\tan^2\delta=\Theta(\log N)$ maximizes $p_s(\delta)$ to $\Theta(1)$, and then $Q_O=Q_\delta$ without any need for amplitude amplification.
It also implies that $Q_O Q_G = \Theta(t/\alpha^2) = \Theta(N\log N)$, which is better than the result of the previous subsection.

For $t=1$ and $\tan^2\delta=\Theta(\log N)$, we recover Tulsi's original algorithm, with $Q_O=Q_G=\Theta(\sqrt{N\log N})$.
For $t=\Theta(\log N)$ and $\tan^2\delta=\Theta(1)$, we obtain $Q_O=\Theta(\sqrt{N})$ and $Q_G=\Theta(\sqrt{N}\log N)$, which is the same computational complexity as in the previous subsection.
The advantage of the multi-step version of Tulsi's algorithm is that $Q_O$ and $Q_G$ can be varied between these two extremes, by an intermediate choice of $t$ while maintaining $t\tan^2\delta=\Theta(\log N)$.
That can be useful when we have to minimize the overall computational complexity of the algorithm, given the effort required to evaluate the oracle as well as the rotation map.

\subsection{Multi-step quantum search on regular graphs:}

Our multi-step algorithm can be used to perform a quantum search on any regular graph.
For the adjacency matrix of general regular graphs, however, we may have an estimate of only the spectral gap $g$, without the knowledge of the entire spectrum.
In such cases, the best known oracle complexity of quantum search is $Q_O=\Theta(\sqrt{\frac{N}{g}})$ for $t=1$ \cite{krovi2016walk,abhijith2018spatial}.
Powering the graph sufficiently many times makes it an expander with $g_t=\Theta(1)$, and then the oracle complexity would attain its optimal scaling behaviour, $Q_O=\Theta(\sqrt{N})$ \cite{ambainis2005coins}.
As per Corollary \ref{cor:1}, our multi-step algorithm replaces $\cos(\phi_k)$ by $\cos^t(\phi_k)$ in the computational complexity analysis.
With $1-g_t = \cos^t(\phi_1) = (1-g)^t \approx e^{-gt}$, we see that $g_t$ would become $\Theta(1)$ for $t=\Theta(\frac{1}{g})$.
Thus our multi-step quantum search algorithm can provide $Q_O=\Theta(\sqrt{N})$ and $Q_G=\Theta(\frac{\sqrt{N}}{g})$ for any regular graph.

The MNRS algorithm \cite{magniez2011walk} achieves the same $Q_O$ with $Q_G=\Theta(\sqrt{\frac{N}{g}})$.
But it requires fresh ancilla qubits at every walk step, and so the total number of qubits required scales as $\Theta(\sqrt{\frac{N}{g}})$.
Compared to this, our graph powering approach requires exponentially fewer number of qubits.
Also, for the specific case of the 2D grid with $g=\Theta(\frac{1}{N})$, the MNRS algorithm is much slower than our result in the previous subsection.
We note, however, that the MNRS algorithm has a wider applicability than our graph powering approach, because it is not limited to quantum walks on regular graphs.

\section{Extension to symmetric Markov chains}
\label{sec:markovchain}

A method to quantize general Markov chains was introduced by Szegedy \cite{szegedy2004quantum}.
In this Section, we describe how to implement a quantum walk corresponding to multiple steps of a classical symmetric Markov chain without explicitly powering the transition matrix of the chain.
We judge the complexity of this procedure in terms of the number of queries made to the elements of the transition matrix.

Let $M = \sum_{i,j} M_{ij} \ket{i}\bra{j}$ be an $N\times N$ stochastic transition matrix of a symmetric Markov chain, i.e $M=M^T$.
Following Szegedy, we define two sets of orthonormal states in $\mathbb{C}^{N^2}$ for $i\in[N]$:
\begin{equation}
\ket{A_i} = \sum_{j=1}^N \sqrt{M_{ij}} \ket{i}\ket{j} ~,~~ \ket{B_i} = \sum_{j=1}^N \sqrt{M_{ij}} \ket{j}\ket{i} ~.
\end{equation} 
We also define two isometric linear operators from $\mathbb{C}^N\rightarrow \mathbb{C}^{N^2}$, using these states:
\begin{equation}
A = \sum_{i=1}^N \ket{A_i}\bra{i} ~,~~ B = \sum_{i=1}^N \ket{B_i}\bra{i} ~,
\end{equation}
and two reflection operators constructed from them:
\begin{equation}
R_1 =  2 A A^\dagger - I ~,~~ R_2 = 2 B B^\dagger - I ~.
\end{equation}
Then the quantum walk corresponding to the Markov chain is defined as,
\begin{equation}
W(M) =  R_2 R_1.
\end{equation}
A single step of this walk is equivalent to two steps of the flip-flop walk \cite{wong2016walk}.

As in the case of flip-flop walks, the walk operator here is also a product of two reflection operators.
If a vector lies in the simultaneous eigenspace of the projectors $A A^\dagger$ and $B B^\dagger$, then $W(M)$ will act on it trivially (either as $I$ or $-I$).
On the other hand, in applications of quantum walks, one is usually interested in subspaces where $W(M)$ acts non-trivially.
It was shown by Szegedy \cite[Theorem 1]{szegedy2004quantum} that the spectral properties of $W(M)$ in the non-trivial subspace are determined by its discriminant matrix,
\begin{equation}
D(M) := A^\dagger B = M ~.
\end{equation}

To implement $W(M)$ one requires operators that construct $\ket{A_i}$ and $\ket{B_i}$ from simpler states.
Let $V_1$ and $V_2$ be the unitary operators that effect this transformation,
\begin{equation}
V_1 \ket{i}\ket{0} = \ket{A_i} ~,~~ V_2 \ket{0}\ket{i} = \ket{B_i}.
\end{equation}
Let the number of queries required to implement each of these transformations be $Q$.
Then a single step of $W(M)$ has the query complexity $4Q$.  

Now we address the question of implementing a quantum walk corresponding to $k$ steps of $M$, i.e. $W(M^k)$.
The naive way to do this is to classically compute the matrix $M^k$, and then quantize it using Szegedy's framework.
This approach has two problems.
First, classical powering of the transition matrix requires us to query all the elements of $M$, with $O(N^2)$ query complexity.
Second, $M$ may have a structure that makes its quantization easy (like random walks on hypercubic lattices), but $M^k$ might not inherit that structure making its quantization difficult. 

We now demonstrate how to implement $W(M^k)$ without classically powering $M$, using techniques very similar to those we used for the flip-flop walk.
To implement $W(M)$, two state registers are required, each having $O(\log N)$ qubits.
To quantize $M^k$, we require $k+1$ such registers.
Let these registers be numbered from $1$ to $k$ starting from the left, and let $V^m_i$ denote $V_i$ applied between the registers numbered $m$ and $m+1$.
Next we define two sets of orthonormal states for $i\in[N]$:
\begin{align}\label{eq:defn_markov}
\ket{A^k_i} = V_1^k \ldots V_1^2 V_1^1 \ket{i}\ket{0} \ldots \ket{0} = \sum_{j_1\ldots j_k} \sqrt{M_{i j_1} M_{j_1 j_2} \ldots M_{j_{k-1} j_k}} \ket{i}\ket{j_1} \ldots \ket{j_k} ~,\\
\ket{B^k_i} = V_2^1 \ldots V_2^{k-1} V_2^k \ket{0} \ldots \ket{0}\ket{i} = \sum_{j_1\ldots j_k} \sqrt{M_{j_{k-1} j_k} \ldots M_{j_1 j_2} M_{i j_1}} \ket{j_k} \ldots \ket{j_1} \ket{i} ~.
\end{align} 
As before, we define two isometric operators from $\mathbb{C}\rightarrow\mathbb{C}^{N^{k+1}}$:
\begin{equation}
A_k = \sum_{i=1}^N \ket{A^k_i}\bra{i} ~,~~ B_k = \sum_{i=1}^N \ket{B^k_i}\bra{i} ~.
\end{equation}
In terms of these operators, we construct a multi-step quantum walk as,
\begin{equation}
W_k(M) = (2 B_k B_k^\dagger - I)(2 A_k A_k^\dagger - I) ~.
\end{equation}

The discriminant of this walk is $A^\dagger_k B_k$.
From the definitions of the states in Eq.\eqref{eq:defn_markov}, it is seen that $A^\dagger_k B_k = M^k$.
So the action of the quantum walk $W_k(M)$ is identical to that of $W(M^k)$ in its non-trivial subspace.
Moreover, each step of $W_k(M)$ requires only $4kQ$ queries to implement.

\section{Discussion}

We have presented an algorithm for quantum search on the two-dimensional grid, which is more efficient compared to AKR and Tulsi's algorithms in terms of the target state oracle complexity $Q_O$.
In terms of the graph structure oracle complexity $Q_G$, it is as good as the AKR algorithm, but is worse than Tulsi's algorithm by a factor of $\sqrt{\log N}$.
Our approach improves the effective connectivity of the graph by taking multiple walk steps inbetween target state oracles.
Naively, one may think that optimization of $Q_O$ would require the graph to be completely connected, as in Grover search, and so the number of walk steps should scale as the diameter of the graph, $O(\sqrt{N})$.
But surprisingly this guess is an overestimate, and we can optimize $Q_O$ by taking only $\Theta(\log N)$ walk steps.

We can understand our result as follows.
The performance of the quantum search algorithm depends, as per Eqs.(\ref{eq:r1}-\ref{eq:r3}), on all the eigenvalues of the graph and how far away they are from $1$.
The naive guess for optimizing the performance is to take enough walk steps to make the graph an expander, i.e. to make sure that all the eigenvalues are at a constant distance away from $1$.
But for the two-dimensional grid, most of the eigenvalues are already far away from $1$, and make constant contributions to the sums in Eqs.(\ref{eq:s1}-\ref{eq:s3}).
The eigenvalues close to $1$, which cause suboptimal behaviour of the algorithm, are comparatively few.
So we need to take only those many number of walk steps, which make the contribution of these few eigenvalues optimal.
As we have shown, $\Theta(\log N)$ walk steps suffice in case of the two-dimensional grid; they increase the spectral gap of the walk operator from $\Theta(\frac{1}{N})$ to only $\Theta(\frac{\log N}{N})$.

Finally, we note that we have improved the effective connnectivity of the graph using the technique of graph powering.
There also exist other graph product techniques, like the replacement product and the zig-zag product, which can improve the connectivity of graphs \cite{reingold2000entropy,hoory2006expander}.
It would be interesting to explore if these techniques can be used to further improve the query complexity of the quantum spatial search problem.

\appendix{: Bounds on relevant sums for the two-dimensional grid}

\noindent
{\bf Bounding $\sum_{\mathbf{k}\ne0} \frac{1}{1-\cos^t\phi_k}$:}
For our algorithm to work well, we need to find a small enough $t$ such that $\sum_{\mathbf{k}\ne0} \frac{1}{1-\cos^t\phi_k} = \Theta(N)$.
For $t=1$, this sum scales as $\Theta(N\log{N})$ \cite{ambainis2005coins}, and we need to improve up on that by choosing a larger $t$.

For any $t$, a simple lower bound for this sum can be derived as follows:
\begin{align}
\sum_{\mathbf{k}\ne0} \frac{1}{1-\cos^t\phi_k} &= \sum_{\mathbf{k}\ne0} \frac{1}{(1-\cos\phi_k)(\sum^{t-1}_{n=0} \cos^n\phi_k)} ~,\\
  &\ge \frac{1}{t}\sum_{\mathbf{k}\ne0} \frac{1}{1-\cos\phi_k} ~,\\
  &= \Omega(\frac{N\log{N}}{t}) ~.
\label{eq:lower_bound}
\end{align}

Deriving a good upper bound requires more work.
For $\theta\in[0,\pi]$, we have
\begin{equation}
\cos\theta \le 1-\frac{2\theta^2}{\pi^2} ~.
\end{equation} 
Using this inequality, together with $1-x \le e^{-x}$, we obtain
\begin{equation}
\cos^t\phi_{\mathbf{k}} = 2^{-t}(\cos(\frac{2\pi k_x}{\sqrt{N}})+\cos(\frac{2\pi k_y}{\sqrt{N}}))^t \le (1-\frac{4\mathbf{k}^2}{N})^t \le e^{-4\mathbf{k}^2 t/N} ~.
\end{equation}
Inserting it in our original sum, we get
\begin{equation}
\sum_{\mathbf{k}\ne0} \frac{1}{1-\cos^t\phi_k} \le \sum_{\mathbf{k}\ne0} \frac{1}{1-e^{-4\mathbf{k}^2 t/N}} ~.
\end{equation} 

Now $\sum_{\mathbf{k}\ne0}$ goes over the points of a two-dimensional grid, with side length $L=\sqrt{N}$.
We can choose $k_i\in\{-\lfloor L/2\rfloor,\ldots,0,\ldots,\lfloor L/2\rfloor\}$, with a weight $1/2$ for the end-points when $L$ is even.
This grid can be divided in to concentric square shells, with the center at the origin, and inner side length $l\in\{1,2,\ldots,\lfloor L/2\rfloor\}$.
The $l^{\rm th}$ shell has $(2l+1)^2-(2l-1)^2 = 8l$ points in it, and for every point,
\begin{equation}
l^2 \leq \mathbf{k^2} \leq 2l^2 ~.
\end{equation}
So for each term on such a square,
\begin{equation}
\frac{1}{1-e^{-4\mathbf{k}^2 t/N}} \le \frac{1}{1-e^{-4l^2 t/N}} ~,
\end{equation}
and adding up all the terms, we get
\begin{equation}
\sum_{\mathbf{k}\ne0} \frac{1}{1-\cos^t\phi_k} \le 8\sum_{l=1}^{\lfloor L/2\rfloor} \frac{l}{1-e^{-4l^2 t/N}} ~.
\label{eq:onedim_sum}
\end{equation}

So far we have managed to upper bound our two-dimensional sum by a simpler looking one-dimensional sum.
To proceed further, we split this sum into two parts at the point $\tilde{l}$, such that $\tilde{l} = \lfloor\sqrt{\frac{N}{4t}}\rfloor$.
Then for all $l\le\tilde{l}$, we ensure that $\frac{4l^2 t}{N}\le1$.
Now for $x\in[0,1]$, we have $1-\frac{x}{2} \ge e^{-x}$, and for $x\ge1$ we have $e^{-x} \le e^{-1}$.
Using these inequalities in the split sum, we obtain
\begin{align}
8\sum_{l=1}^{\lfloor L/2\rfloor} \frac{l}{1-e^{-4l^2 t/N}} &\le \frac{4N}{t}\sum_{l=1}^{\tilde{l}} \frac{1}{l} + \frac{8}{1-e^{-1}}\sum_{l=\tilde{l}+1}^{\lfloor L/2\rfloor} l ~,\\
 &= O(\frac{N}{t}\log(\tilde{l}) + L^2) ~,\\
 &= O(\frac{N}{t}\log{\frac{N}{t}} + N) ~.
\label{eq:upper_bound}
\end{align}

We can bring together the lower bound of Eq.\eqref{eq:lower_bound}, and the upper bound of Eq.\eqref{eq:upper_bound}, by choosing $t=O(\log N)$.
Then we get the result, $\sum_{\mathbf{k}\neq0} \frac{1}{1-\cos^t\phi_k} = \Theta(\frac{N\log N}{t})$.  
Our optimal choice $t=\Theta(\log N)$ gives, $\sum_{\mathbf{k}\neq0} \frac{1}{1-\cos^t\phi_k} = \Theta(N)$.
   
\smallskip\noindent
{\bf Bounding $\sum_{\mathbf{k}\ne0} \frac{1}{(1-\cos^t\phi_k)^2}$:}
We can bound this sum also by repeating the same steps.
For $t=1$, this sum scales as $\Theta(N^2)$ \cite{ambainis2005coins}, and we want to improve up on that.
A lower bound can be found, in a manner analogous to the derivation of Eq.\eqref{eq:lower_bound}, as
\begin{equation}
\sum_{\mathbf{k}\ne0} \frac{1}{(1-\cos^t\phi_k)^2} = \Omega(\frac{N^2}{t^2}) ~.
\label{eq:lower_bound2}
\end{equation}

To find the upper bound, first we use the same steps that led to Eq.\eqref{eq:onedim_sum}, converting the two-dimensional sum to a one-dimensional one,
\begin{equation}
\sum_{\mathbf{k}\ne0} \frac{1}{(1-\cos^t\phi_k)^2} \le \sum_{\mathbf{k}\ne0} \frac{1}{(1-e^{-4\mathbf{k}^2 t/N})^2} \le 8\sum_{l=1}^{\lfloor L/2\rfloor} \frac{l}{(1-e^{-4l^2 t/N})^2} ~.
\end{equation} 
We then split the one-dimensional sum in to two parts at $\tilde{l}=\lfloor\sqrt{\frac{N}{4t}}\rfloor$, as before.
Working out the details, we find
\begin{align}
8\sum_{l=1}^{\lfloor L/2\rfloor} \frac{l}{(1-e^{-4l^2 t/N})^2} &\le \frac{2N^2}{t^2}\sum_{l=1}^{\tilde{l}} \frac{1}{l^3} + \frac{8}{(1-e^{-1})^2}\sum_{l=\tilde{l}+1}^{\lfloor L/2\rfloor} l ~,\\
 &= O(\frac{N^2}{t^2} + N) ~.
\label{eq:upper_bound2}
\end{align}
For any $t=O(\sqrt{N})$, the lower bound of Eq.\eqref{eq:lower_bound2} and the upper bound of Eq.\eqref{eq:upper_bound2} are brought together, giving $\sum_{\mathbf{k}\ne0} \frac{1}{(1-\cos^t\phi_k)^2} = \Theta(\frac{N^2}{t^2})$.

\smallskip\noindent
{\bf Bounding $\sum_{k\ne0}\cot^2(\phi^{(t)}_k/2)$:}
Using $\cot^2\theta = \text{cosec}^2\theta - 1$, and Corollary \ref{cor:1}, we have
\begin{equation}
\sum_{k\ne0}\cot^2\frac{\phi^{(t)}_k}{2} = 1-N + \sum_{k\ne0}\text{cosec}^2\frac{\phi^{(t)}_k}{2} = 1-N + \sum_{k\ne0} \frac{2}{1-\cos^t\phi_k} ~.
\end{equation}
Hence, Eq.\eqref{eq:lower_bound} implies that $\sum_{k\ne0}\cot^2(\phi^{(t)}_k/2)$ has the lower bound $\Omega(\frac{N}{t}\log N)$, and Eq.\eqref{eq:upper_bound} implies that it has the upper bound $O(\frac{N}{t}\log\frac{N}{t} + N)$.
The two bounds come together for the choice $t=O(\log N)$, giving $\sum_{k\ne0}\cot^2({\phi}^{(t)}_k/2) = \Theta(\frac{N\log N}{t})$.
Our optimal choice $t=\Theta(\log N)$ gives, $\sum_{k\ne0}\cot^2({\phi}^{(t)}_k/2) = \Theta(N)$.

\appendix{: Proof that $\alpha<\frac{1}{2}\phi^{(t)}_1$}

\noindent
As per Corollary \ref{cor:1}, we have $\cos\phi^{(t)}_1 = \cos^t\phi_1$.
Our strategy is to lower bound $\phi^{(t)}_1$ with a quantity asymptotically greater than $\alpha$.
To this end, we use the inequalities,
\begin{equation}
1-\frac{\theta^2}{2} \le \cos\theta \le 1-\frac{2\theta^2}{\pi^2} ~, 
\end{equation}
for $\theta\in[0,\pi]$, to express
\begin{equation}
\phi^{(t)}_1 \ge \sqrt{2(1-\cos^t\phi_1)} \ge \sqrt{2\left(1-(1-\frac{2\phi^2_1}{\pi^2})^t\right)} ~.
\end{equation}
Then using the inequality $1-x\le e^{-x}$, we have the bound,
\begin{equation}
\phi^{(t)}_1 \ge \sqrt{2\left( 1-e^{-2\phi^2_1 t/\pi^2} \right)} ~.
\end{equation}

For the two-dimensional grid, $\cos\phi_1 = \cos^2\frac{\pi}{\sqrt{N}}$ and $\phi_1 = \Theta(\frac{1}{\sqrt{N}})$.
So for the choice $t=\Theta(\log N)$, the exponent $\frac{2\phi^2_1 t}{\pi^2}$ is much smaller than $1$.
Combining this fact with the inequality $1-\frac{x}{2}\ge e^{-x}$ for $x\in[0,1]$, we finally get
\begin{equation}
\phi^{(t)}_1 \ge \sqrt{\frac{2\phi^2_1 t}{\pi^2}} = \Omega(\sqrt{\frac{t}{N}}) ~.
\end{equation}
From Eq.\eqref{eq:s1}, we know that $\alpha=\Theta(\sqrt{\frac{t}{N\log N}})$.
As a result, the condition $\alpha<\frac{1}{2}\phi^{(t)}_1$ holds asymptotically.

\end{document}